\documentclass[reprint]{revtex4-1}
\usepackage{graphicx}
\usepackage{amsmath,amssymb,amsthm,mathtools}
\usepackage{bm}
\usepackage{setspace}
\usepackage{hyperref}
\usepackage[dvipsnames]{xcolor}
\DeclareMathOperator{\arccsc}{arccsc}

\begin{document}

\title{Synchronization on star-like graphs and emerging \texorpdfstring{$\mathbb{Z}_{p}$}{Z(p)} symmetries at strong coupling}

\author{Artem Alexandrov}
\affiliation{Moscow Institute for Physics and Technology, Dolgoprudny 141700, Russia}

\author{Pavel Arkhipov}
\affiliation{Moscow Institute for Physics and Technology, Dolgoprudny 141700, Russia}

\author{Alexander Gorsky}
\affiliation{Institute for Information Transmission Problems RAS, 127051 Moscow, Russia}
\affiliation{
Moscow Institute for Physics and Technology, Dolgoprudny 141700, Russia}

\begin{abstract}
We discuss the aspects of  synchronization on  inhomogeneous star-like graphs with long rays in Kuramoto model framework. We assume the positive correlation between  internal frequencies and  degrees for all nodes which supports the abrupt first order synchronization phase transition. It is found that different ingredients of the graph get synchronized at different critical couplings. Combining numerical and analytic tools we evaluate all critical couplings for the long star graph. Surprisingly it is found that at strong coupling  there are discrete values of coupling constant which support the synchronized states with emerging $\mathbb{Z}_{p}$ symmetries. The stability of synchronized phase is discussed and the interpretation of phase with emerging $\mathbb{Z}_{p}$ symmetry for the Josephson array on long star graph is mentioned.
\end{abstract}

\maketitle

\section{Introduction}

Synchronization phase transition is the phenomenon  responsible for a collective behavior of many-body systems in many physical, social and biological problems. Usually it was assumed that the phase transition at some critical coupling is of the second order (see \cite{Arenas2008Review} for the review) however more recently the pattern of  abrupt first order phase transition has been recognized \cite{Arenas2019Review}. The simplest model governing the synchronization has been suggested by Kuramoto \cite{Kuramoto1984}  and the big family of models generalizing the initial version has been suggested later. The list of various modifications can be found in \cite{Arenas2008Review,Arenas2019Review}.

In the initial Kuramoto model one considers the degrees of freedom living at the nodes of the full graph that is the system involves all-to-all interactions. Such simple model with all-to-all interaction  can be  naturally generalized for a arbitrary graph whose adjacency matrix encodes more complicated interaction pattern. It was found that some aspects of the 
graph architecture strongly influence the synchronization pattern however the exact impact of the different characteristics of the graph adjacency matrix on the criticality in the Kuramoto model was not elaborated enough. Only several simplest graphs like circle \cite{Ochab2009,Roy2012}, star \cite{Arenas2011ExplosiveSync,Vlasov2015} or two-star graph \cite{Wang2017,Chen2017} have been analysed in details. For instance, when the degrees of the nodes in graph and corresponding internal frequencies are equal the abrupt synchronization for the star graph occurs \cite{Arenas2011ExplosiveSync,Vlasov2015}. Different aspects of synchronization on  star graphs have been investigated in \cite{Arenas2011ExplosiveSync,Boccaletti2019,borisyuk}.

To some extend the  synchronization phenomenon is the classical counterpart of the Bose-Einstein condensation (BEC) of charged degrees of freedom hence one could question about the properties of BEC on  graphs under consideration. Indeed in was found \cite{Burioni2000,Burioni2002,current} that the BEC depends on the topological graph properties and on the star graphs condensate becomes inhomogeneous decaying at  distance from the hub. The condensate gets formed by effective filling of the negative mode which exists in the graph spectrum. The phenomenon has been investigated experimentally for the Josephson junctions organized as star graph arrays \cite{Lorenzo2014} and the expected inhomogeneous transport phenomena have been found. More recently the similar effects have been found for tree graphs \cite{tree}.

There are some important symmetry issues underlying the Kuramoto dynamics for 
simplest graphs. For instance for the full graph the Kuramoto dynamics with $N$ nodes enjoys $(N-3)$ conservation laws \cite{Watanabe1993,Watanabe1994} and the whole dynamics gets reduced to the geodesic motion on the M\"{o}bius  group. The integrals of motion get identified with the independent cross-ratios of $N$ points on $S^1$. The  M\"{o}bius group plays the key role in the derivation of the critical couplings for the Kuramoto model on the star graph as well however the role of the M\"{o}bius group for dynamics on more general graphs has not been clarified yet.

In this note we shall investigate Kuramoto model on more complicated star-like graphs namely ``long star'', ``decorated long star'' and ``neuron''-like graph. Our aim is to elucidate the critical behavior in all cases identifying the critical couplings and the order of a phase transition. We assume a positive correlation of  degrees and  internal frequencies for all nodes (see \cite{Dorog} for the discussion concerning this point). We have found a step-like synchronization pattern when hub, rays and leaves get synchronized at the different critical couplings and the clear-cut hysteresis for the first order transitions is seen numerically. Similar step-like synchronization has been observed for two-star graph \cite{Wang2017}. At intermediate values of coupling we have found a kind of chimera state when part of degrees of freedom gets synchronized while the rest is in desynchronized state. Such step-like synchronization pattern has been found for all three types of graphs.

The quite unexpected phenomenon we found concerns the naively totally
synchronized phase at large Kuramoto coupling. It turns out that for long star graph and for decorated long  star graph at large number of rays there are discrete values of coupling constant  when the standard Kuramoto order parameter vanishes. The number of such points is finite and proportional to the number of nodes on the rays. We argue that at these points the system is 
fully synchronized at strong coupling  with emerging $\mathbb{Z}_k$-symmetry state for some $k$ which depends on the value of the coupling constant. The large $N$ case can be treated analytically and the numerical simulations confirm the existence of such values of couplings with enhanced symmetry.

The paper is organized as follows. In the \hyperref[sec:MultiStepSync]{Section 2} we provide some details concerning  synchronization phenomena on graphs,   describe our setup and perform numerical simulations. The multistep synchronization process and hysteresis phenomena for the Kuramoto model on the long star, decorated star graphs and ''neuron'' graph are described. 
In \hyperref[sec:EmergingZSymm]{Section 3} we explain how $\mathbb{Z}_k$-symmetry appears and present analytical and numerical results at strong coupling. In the \hyperref[sec:Discussion]{Discussion}  we summarize our findings and discuss possible relations of observed phenomena to Josephson arrays on the graphs.

\section{Multi-step synchronization on star-like graphs}
\label{sec:MultiStepSync}
\subsection{General remarks}
The Kuramoto model on the generic graph is described by the 
equations
\begin{equation}\label{general-Kuramoto}
        \dot{\theta}_i = \omega_i + \lambda \sum_{j=1}^N A_{ij} \sin\left(\theta_j - \theta_i\right),
        \end{equation}
where $A$ is adjacency matrix of the graph, $\lambda$ is the coupling constant,  $\omega_i$ are oscillator frequencies, $N$ is the total number of nodes. In this paper we  consider the case  $\omega_i=\sum_jA_{ij}=d_i$ where $d_i$ is degree of $i$-th node. This choice for the star graph yields the first order phase transition with hysteresis \cite{Arenas2011ExplosiveSync}. The order parameter for the Kuramoto model is defined as $R(\lambda)=\langle|\sum_ke^{i \theta_k}/N|\rangle$ where time averaging is assumed,
\begin{equation}
    R(\lambda)=\left\langle\left|\frac{1}{N}\sum_ke^{i \theta_k}\right|\right\rangle = \lim\limits_{T\rightarrow\infty}\frac{1}{T}\int_{0}^{T}\frac{dt}{N}\left|\sum_{i=k}^{N}e^{i\theta_k}\right|
\end{equation}
The case $R\rightarrow 1$ corresponds to the complete synchronization while $R\rightarrow 0$ to the desynchronized state.

In pioneering works \cite{Watanabe1994} it was shown that Kuramoto model on a full graph with equal eigenfrequencies exhibits low-dimensional dynamics, that is for case of graph with $N$ vertices the number of degrees of freedom can be reduced with help of $N-3$ integrals of motion. In  \cite{Chen2017,Marverl2009} the authors have proposed elegant and rigorous description of such integrals of motion  and have showed that low-dimensional dynamics of model gets reduced to the flow on M\"{o}bius group. The major statement of these works sounds as follows: if the model can be rewritten in the following way,
\begin{equation}\label{reducible_models}
    \dot{\theta}_i = g + fe^{i\theta_i} + \bar{f}e^{-i\theta_i},
\end{equation}
where the functions $f$ and $g$ depend only on time, the low-dimensional dynamics  can be described as the flow on the orbit of M\"{o}bius group. 

\begin{figure}
\begin{minipage}{0.49\linewidth}
    \centering
    \includegraphics[width=\linewidth]{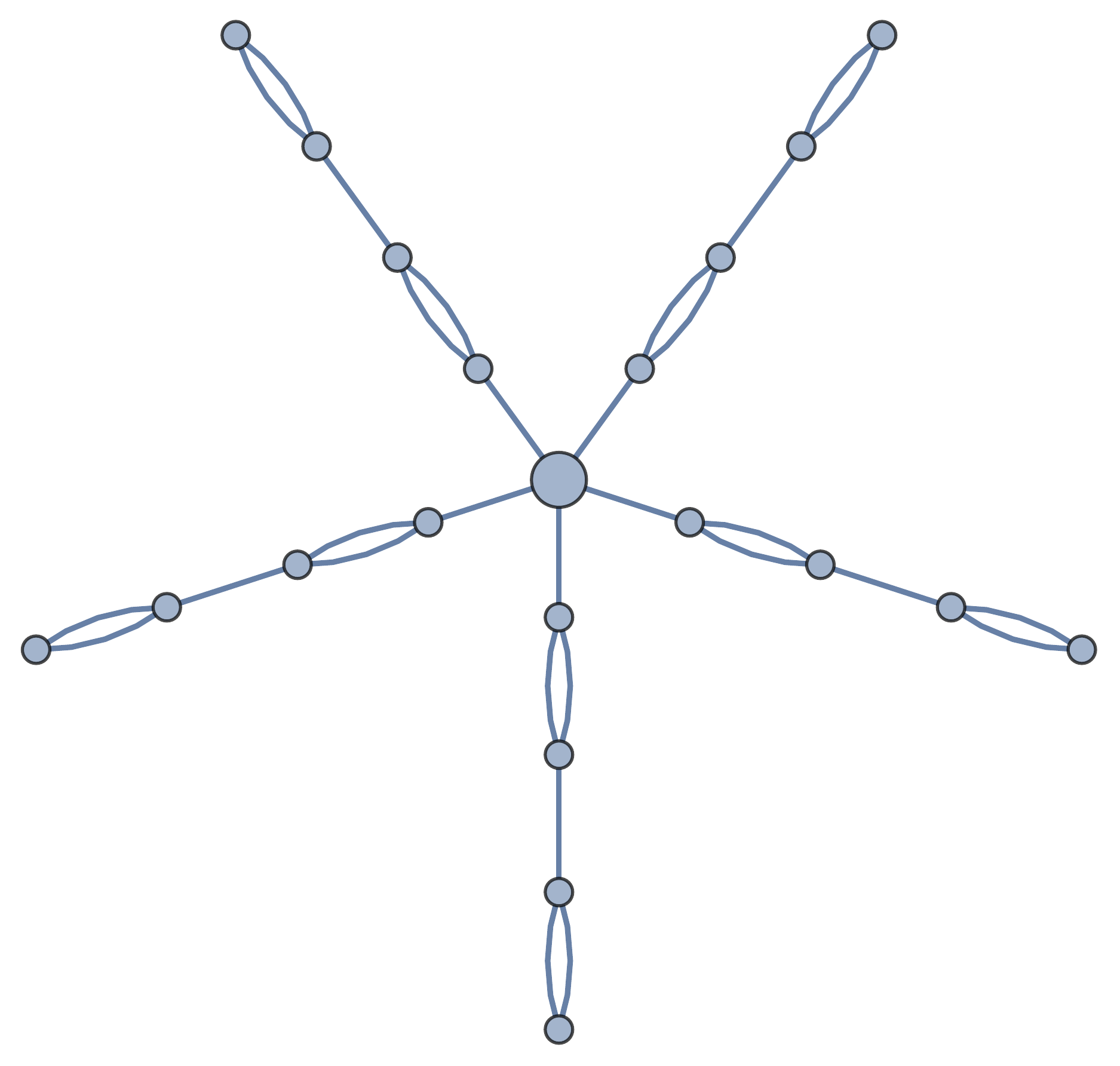} \\ a)
\end{minipage}
\hfill
\begin{minipage}{0.49\linewidth}
    \centering
    \includegraphics[width=\linewidth]{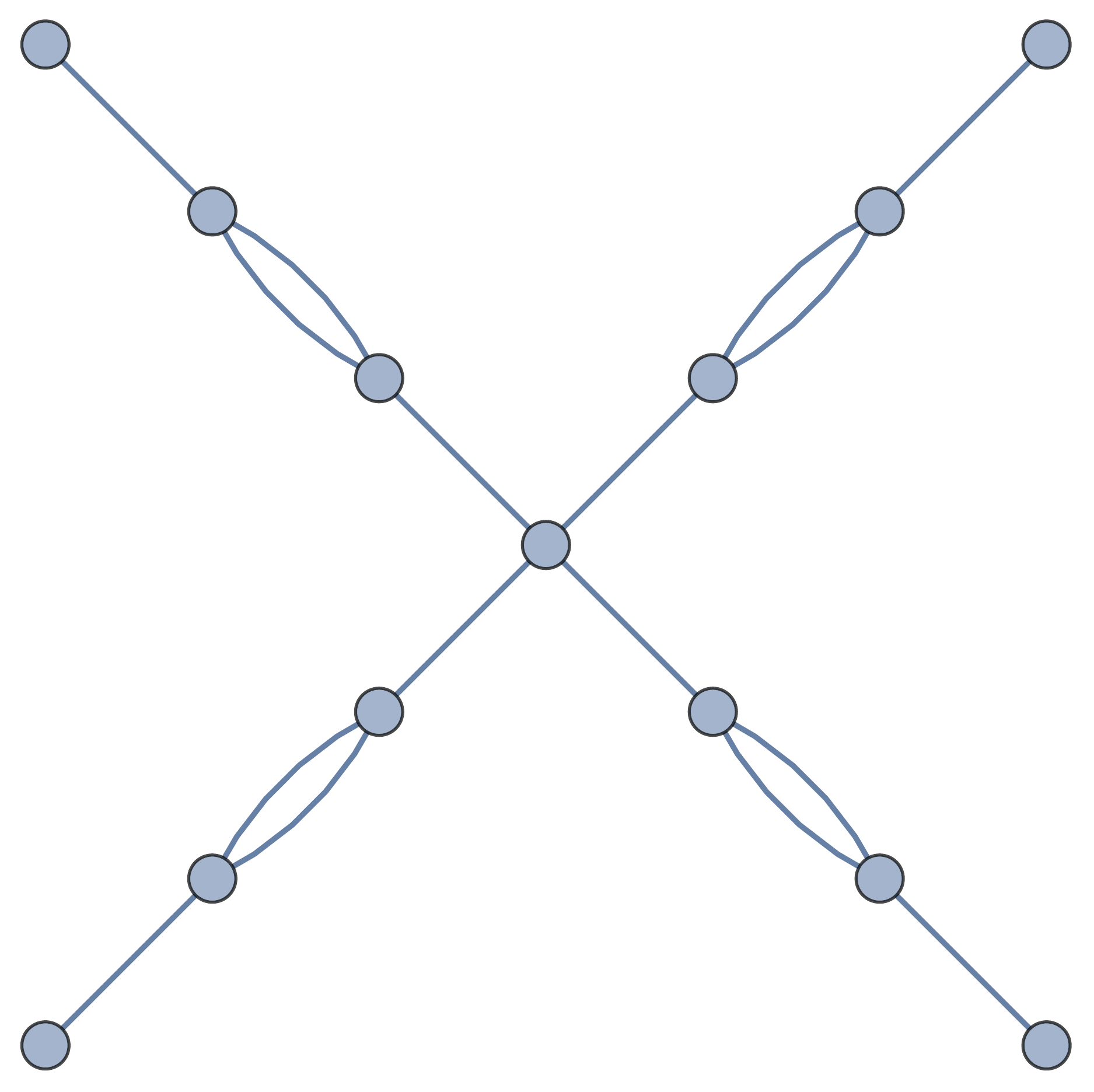} \\ b)
\end{minipage}
    \caption{a) decorated star graph with $n=5$ rays, $p=4$ ray length and $m=2$ additional bonds on the ray; b) $n=4$, $p=2$, $m=2$}
    \label{fig:dstar_examples}
\end{figure}

\begin{figure}
    \centering
    \includegraphics[width=\linewidth]{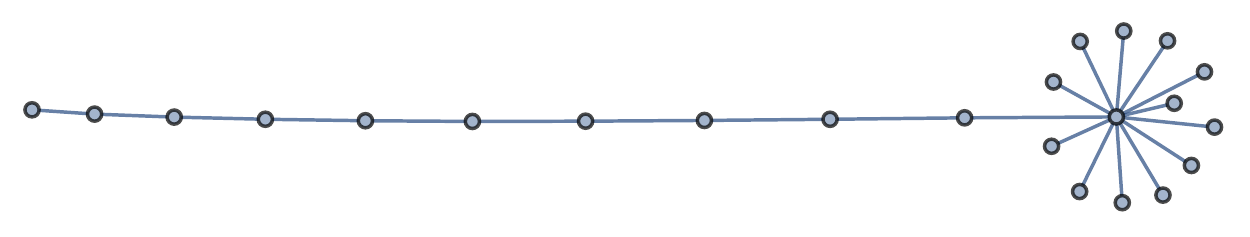}
    \caption{``Neuron'' graph with $n=12$ rays and tail of $p=10$ length}
    \label{fig:neuron_example}
\end{figure}

Later in the paper \cite{Vlasov2015} Kuramoto model was considered on the star graph. The model on such graph demonstrates the first order phase transition with respect to coupling constant $\lambda$. The authors have noticed that rewriting  original model in terms of phase differences between hub and $i$-th leave, $\psi_i$  yields exact values of critical constants in the limit of large number of leaves. This observation is closely related to the statement concerning dimensional reduction. Indeed, the system of equations for phase differences $\psi_i$ can be rewritten in form \eqref{reducible_models} with
\begin{equation}
    f=\frac{i\lambda}{2},\quad g= -(\beta-1)\omega-\frac{\lambda\beta}{N}\sum_{j=1}^{N}\sin\psi_j,
\end{equation}
where $N$ is the number of rays, $\omega$ \& $\beta\omega$ are eigenfrequencies of the each leave and the hub respectively ($\beta>0$). Finally, in the work \cite{Boccaletti2019} a stability of synchronized phase was discussed in details.
\begin{figure}
\begin{minipage}{0.49\linewidth}
    \centering
    \includegraphics[width=\linewidth]{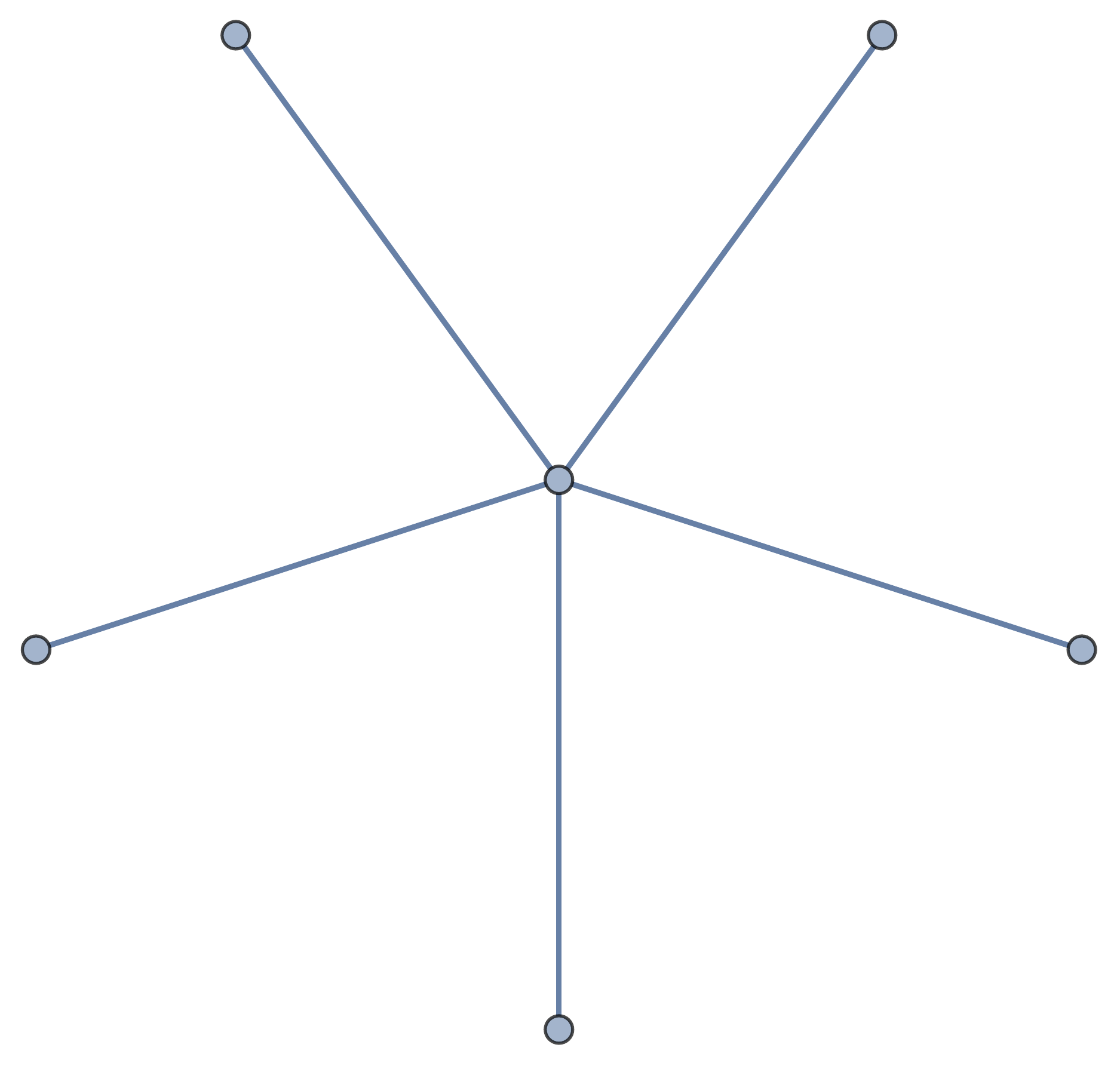} \\ a)
\end{minipage}
\hfill
\begin{minipage}{0.49\linewidth}
    \centering
    \includegraphics[width=\linewidth]{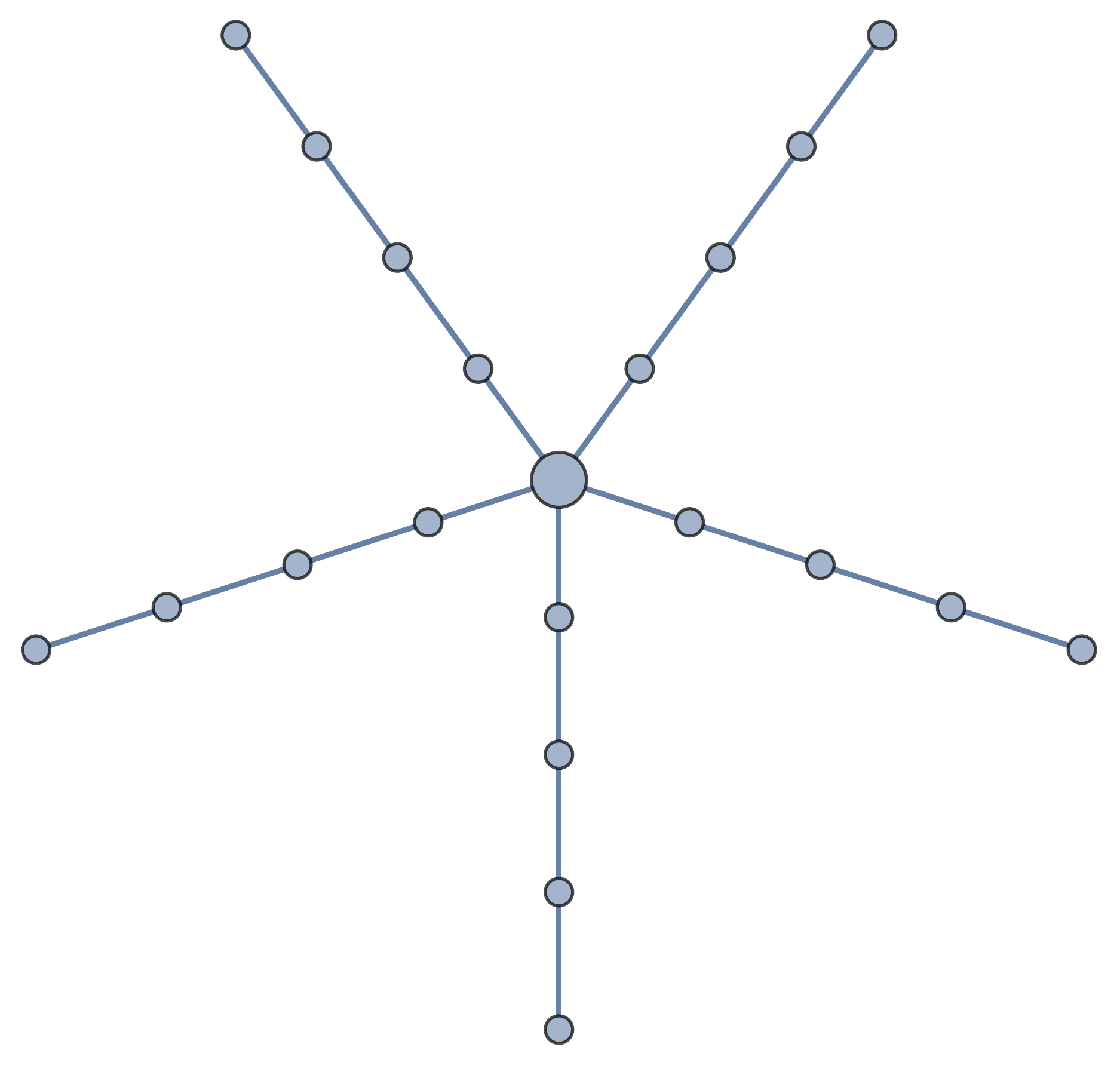} \\ b)
\end{minipage}
    \caption{a) star graph with $n=5$ rays; b) long star graph with $n=5$ rays and ray length $p=4$}
    \label{fig:star_lstar}
\end{figure}
In this note we investigate Kuramoto model on three types of graphs. First, we consider the model on a star with long rays, which we denote as ``long star graph'' (see fig.~\ref{fig:star_lstar}): it is a star graph with $n$ rays and $p$ additional nodes on each ray. Second, we consider the model on a star with long rays and additional $m$ bonds through the ray ($m$ additional bonds on each even edge), denoted as ``decorated star graph'' (see fig.~\ref{fig:dstar_examples}) which can be thought of as the long star with some weights for links. The third type of graph considered is the ``neuron'' -- like long star graph with one long ray (see fig.~\ref{fig:neuron_example}).
We analyze the order parameter behavior $R=R(\lambda)$ for such types of graphs and find some interesting features concerning hysteresis in the Kuramoto model and the synchronization process dynamics.

\subsection{Long star graph}

\begin{figure}
    \centering
    \includegraphics[width=0.9\linewidth]{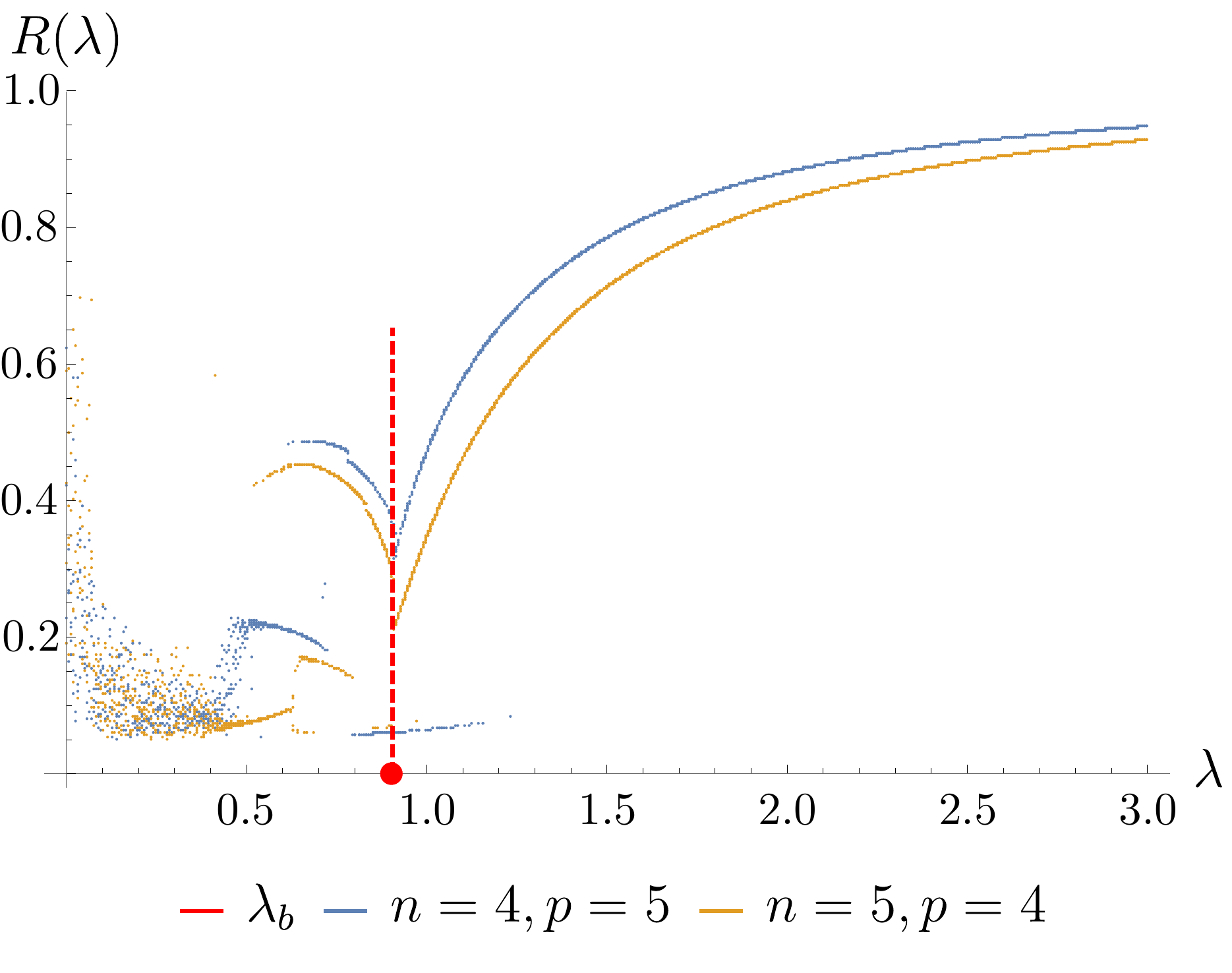} \\ b)
\caption{Order parameter $R=R(\lambda)$ for the model on long star graph with a) fixed $n=5$ and different values of $p$; b) with fixed quantity $n\cdot p$}
\label{fig:lstar_order}
\end{figure}

\begin{figure}
    \centering
    \includegraphics[width=0.8\linewidth]{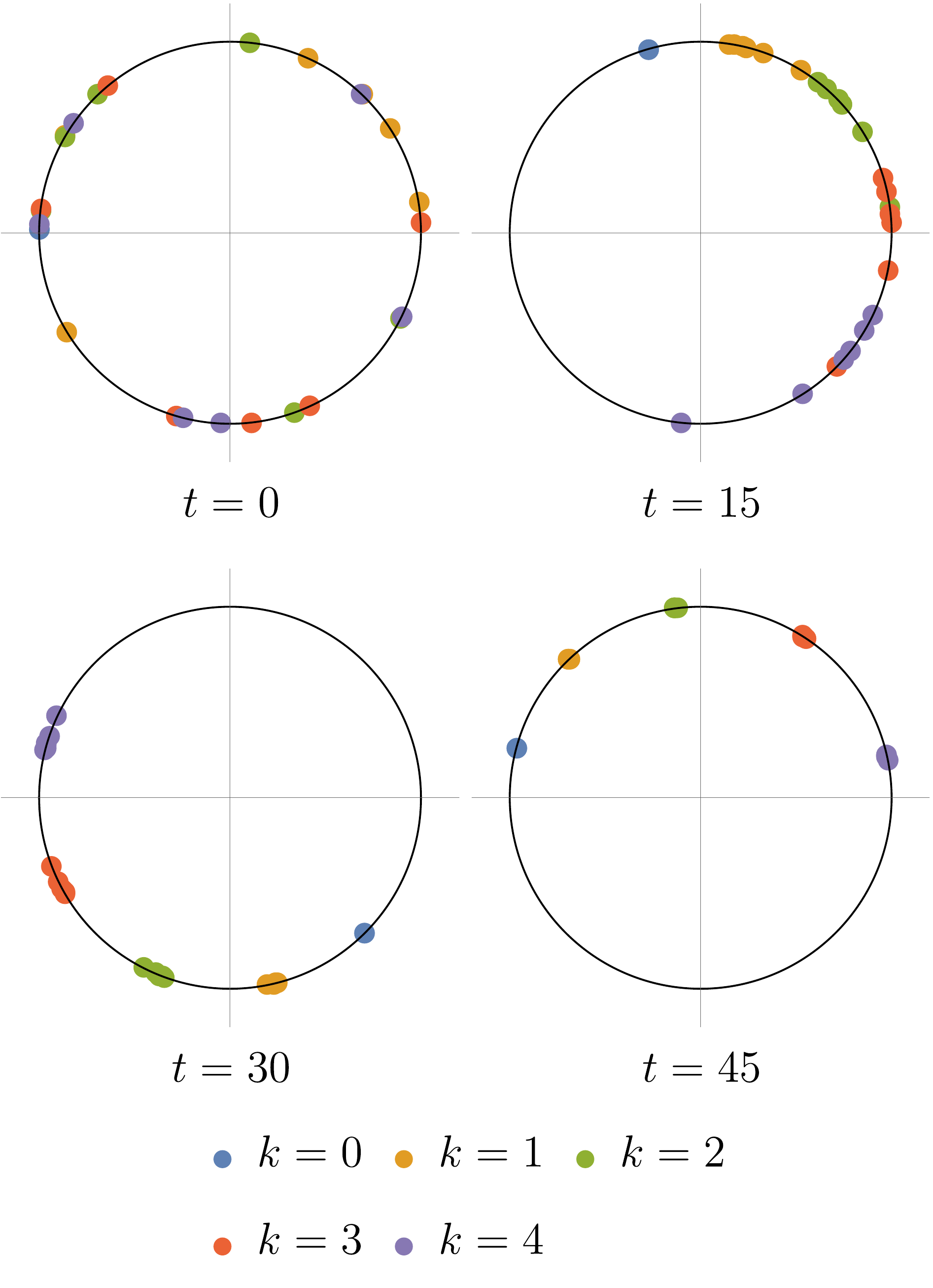}
    \caption{Synchronization dynamics on the long star with $n=6$, $p=4$, the index $k$ denotes the layer, $k=0$ represents the hub}
    \label{fig:lstar_snapshots}
\end{figure}

We start our consideration with numerical simulation in Kuramoto model on the long star graph. In order to investigate the full phase space, we implement the following numeric simulation scheme:
\begin{flushleft}
    \setstretch{0.3}
    \noindent\rule{\linewidth}{0.6pt}\\
    Simulation scheme \\
    \noindent\rule{\linewidth}{0.6pt} \\
    {\bf set} $\lambda_{\max}$ \\
    {\bf while} $\lambda <\lambda_{\max}$:\\
    \begin{enumerate}
        \itemsep0em 
        \item choose uniformly distributed random initial conditions $\theta_i(t=0)\in [0,2\pi)$
        \item find the solution $\theta_i=\theta_i(t)$ for $t\in [0,T]$ for the given value of coupling constant $\lambda_0$, where $T$ is large
        \item compute the order parameter $R=R(\lambda_0)$ by numerical integration
        \item increase the coupling constant $\lambda_0$ by the small quantity $\delta \lambda$
    \end{enumerate}
    {\bf end}
\end{flushleft}
In such scheme we vary the coupling constant quasistatically because we compute average over large time interval. For a given value of coupling constant $\lambda$, we choose uniformly distributed random initial conditions on interval $[0,2\pi)$ and perform numerical solution of the system \eqref{general-Kuramoto} with adjacency matrix $A_{ij}$ corresponding to long-star graph. Having obtained set of numerical solutions $\theta_i(t)$, we perform averaging over large interval $T$. It is found that the order parameter curve $R=R(\lambda)$ has three critical points, which can be recognized from the fig.~\ref{fig:lstar_order}.
\begin{figure}
    \centering
    \includegraphics[width=0.9\linewidth]{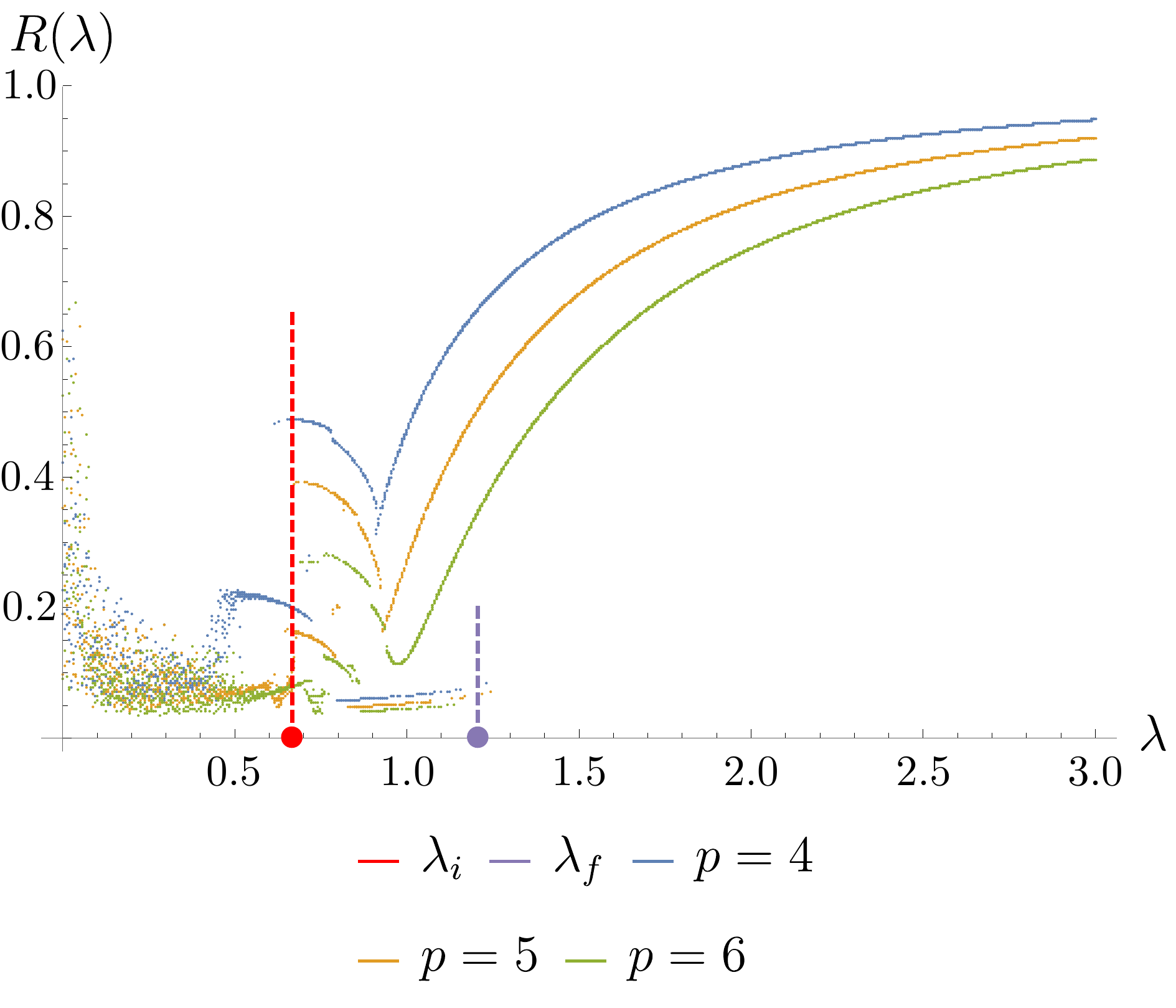} \\ a)
\end{figure}
We see that hysteresis appears for the long star graph as similarly to the case of  simple star graph \cite{Vlasov2015}. During the synchronization process the oscillators form $N_c=p+1$ clusters. From the system snapshots (see fig.~\ref{fig:lstar_snapshots}) (we describe dynamics in terms of complex phases $z_i=\exp(i\theta_i)$, living on unit disk in complex plane) at different times, we can conclude that the synchronization starts from the hub and propagates through rays. These empirical observations are in agreement with the curve $R=R(\lambda)$. 
Numerical simulations with different parameters of graph (variations of the rays number $n$ and the ray length $p$) shows that two of three critical points are the same as for star graph,
\begin{equation}
    \lambda_c^{i}=\frac{n-1}{n+1},\quad\lambda_c^{f} = \frac{n-1}{\sqrt{2n+1}}.
\end{equation}
This fact reaffirms empirical observation concerning the early times of synchronization process. We argue that the value of  $N_c$ allows the simple explanation: it just tells us that there are $p$ clusters corresponding to the length of ray and there is one cluster with the hub only. We observe that there is a phase difference between the oscillators on the same ray. Also, in works \cite{Vlasov2015} the authors discussed the synchronization process on the star graph and from the system snapshots one can see the phase difference between the hub and leaves appears as well. 

The last critical point can be  find analytically. Indeed, one can consider the system in the synchronized state and start to decrease coupling constant $\lambda$. In the fully synchronized state, all the oscillators have the average frequency $\Omega = 2np/(np+1)$ and the general solution for $i$-th oscillator is simply $\theta_i(t)=\Omega t + \varphi_i$, where $\varphi_i$ is a constant. Substituting this solution into the equations of motion we obtain 
\begin{equation}
    \Omega = \omega_i + \lambda\sum_{j=1}^{N}A_{ij}\sin\left(\varphi_j-\varphi_i\right).
\end{equation}
In the synchronized state all the rays of the long star graph are symmetrical. Using basic algebra, one can show that the greatest value among $\left|\lambda \sin (\theta_k - \theta_{k+1})\right|$, where the index $k$ denotes the distance from the hub, is $(np - 1) / (np + 1)$, for $k=p$. That means, if we start to decrease the coupling constant, the synchronized solution will stop existing when $\lambda$ becomes less than $(np - 1) / (np + 1)$, as $\sin(\theta_p - \theta_{p+1}) \leq 1$. For the behaviour of the system, this would mean that the leaves get desynchronized first. The final expression for the backward critical coupling constant is
\begin{equation}
    \lambda_c^{b}=\frac{np-1}{np+1}.
\end{equation}
Therefore numerical simulations supplemented with the analytic arguments allow us to find out all critical constants for long star graph. Having obtained all critical constants, it is possible to implement well-known simulation scheme, where a solution found at $k$-th step is used as the initial condition at $(k+1)$-th step (for instance, see \cite{Vlasov2015}). In this scheme hysteresis is clearly observable.

Now we discuss stability of the synchronized state. The corresponding weakly perturbed solution can be represented as $\theta_{i}(t)=\Omega t + \varphi_i + \delta\theta_i(t)$, where $\varphi_i$ is constant and $\delta\theta_i(t)$ is a small perturbation over the synchronized state. Substituting this solution into the equations of motion, we obtain
\begin{equation}
    \delta\dot{\theta}_i + \Omega = \omega_i + \lambda\sum_{j=1}^{N}A_{ij}\cos\left(\varphi_j-\varphi_i\right)\left(\delta\theta_j - \delta\theta_i\right),
\end{equation}
in the linearized approximation
\begin{equation}\label{linear_approx}
    \delta\dot{\theta}_i= \lambda \sum_{j=1}^{N}A_{ij}\Delta_{ji}\left(\delta\theta_j - \delta\theta_i\right),\Delta_{ji} = \cos\left(\varphi_j-\varphi_i\right).
\end{equation}
Stability of the system can be analyzed with Gershgorin theorem. Handling with indices, one can rewrite \eqref{linear_approx} as $\delta\dot{\theta}_i = -\lambda T_{ik}\delta\theta_k$, where $T_{ik}$ is the square matrix and we assume summation over $k$. According to Gershgorin theorem \cite{Ochab2009}, the eigenvalues of matrix $T_{ij}$ are localized on circles of radius $R_i=\sum_{k=1}T_{ik}$ with centers at $T_{ii}$. The matrix $T_{ik}$ has a zero eigenvalue, which corresponds to the homogeneous translations, $\theta_i\rightarrow\theta_i+\chi$, where $\chi$ is the constant (see \cite{Ochab2009} and \cite{Boccaletti2019} for more details). Simple analysis tells us that if $\forall i,j\in\lbrace 1,...,N\rbrace\rightarrow|\varphi_j-\varphi_i|<\pi/2$ then all the eigenvalues of $T_{ij}$ are non-negative, and the synchronized solution is stable. Similarly, if $|\varphi_j-\varphi_i|>\pi/2$ the solution is unstable. If one find $i$ and $j$ that $|\varphi_j-\varphi_i|>\pi/2$ and also $|\varphi_j-\varphi_i|<\pi/2$ appears, the more sophisticated analysis is needed.

The intermediate regime when only part of ingredients, say rays, get synchronized can be considered as an example of chimera states discussed in context of Kuramoto model (see \cite{Abrams2006}, \cite{Laing2009} for discussion of chimera states). In \cite{Abrams2008} the author showed the existence of chimera state in case of two identical all-to-all coupled oscillators populations and in \cite{Pikovsky2008} authors provided more detailed discussion of this system. The case of long star is in some sense similar: this graph has non-trivial adjacency matrix with
additional symmetry.  
	
\subsection{Decorated star graph}

\begin{figure}
    \centering
    \includegraphics[width=0.8\linewidth]{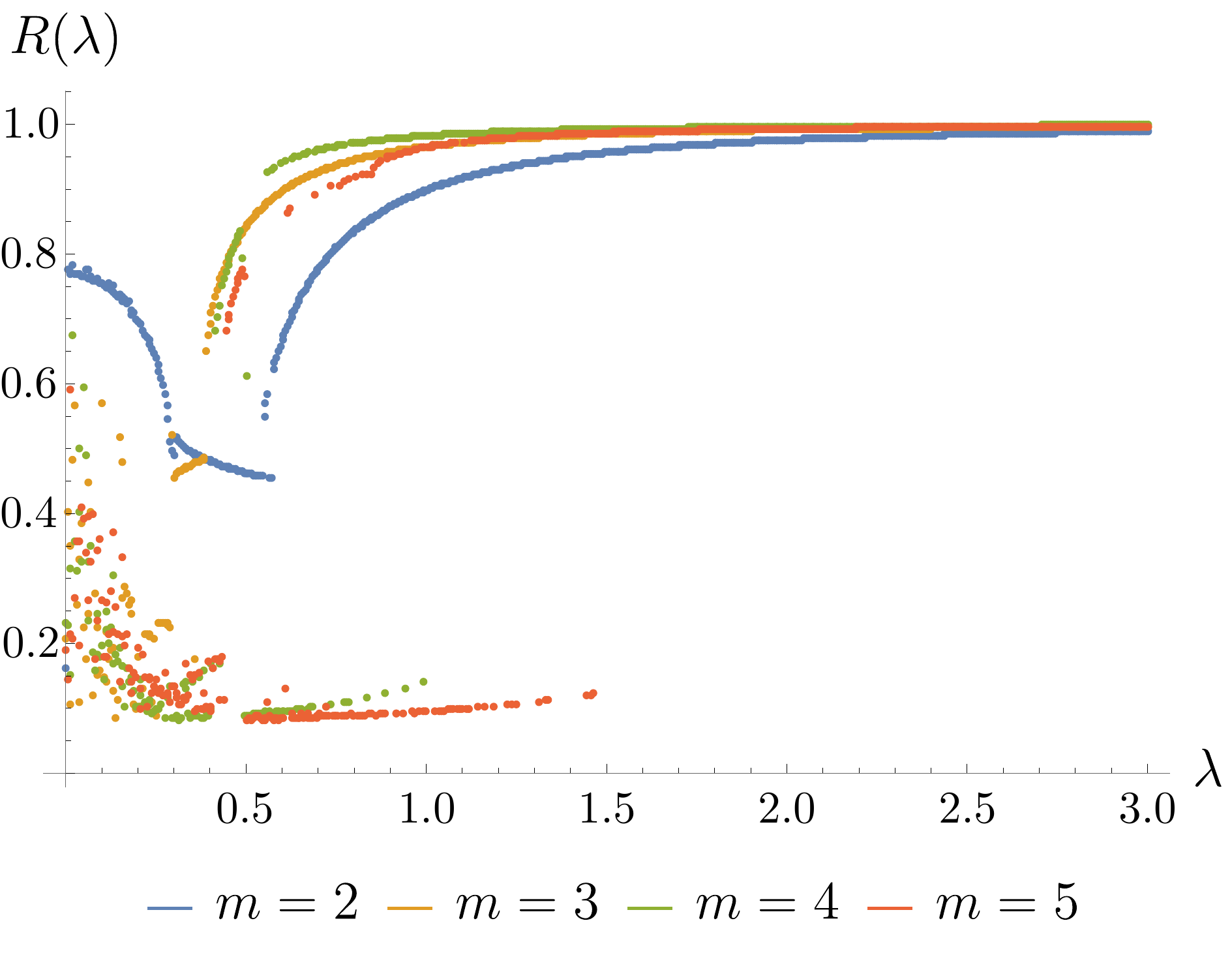} \\ b)
\caption{a) Comparison of phase diagrams for long star graph with $n=3$ \& $p=4$ and decorated star with $n=3$, $p=4$ \& $m=2$; b) Phase diagrams for decorated star with $n=3$, $p=4$ and different $m$}
\label{fig:compare_lstar_dstar}
\end{figure}

Consider now modification of long star graph -- ``decorated star''. The decorated star graph is defined by three quantities: the number of rays $n$, the length of each ray $p$ and the number of additional edges on each ray $m$. It is clear that for $m=1$ the decorated star graph is just the long star graph. So, it is reasonable to expect that for small values of $m$ the phase diagram of the Kuramoto model on the decorated star graph is quite similar to the corresponding phase diagram of the long star graph. It can be seen from the following plots (fig.~\ref{fig:compare_lstar_dstar}a).
\begin{figure}
    \centering
    \includegraphics[width=0.8\linewidth]{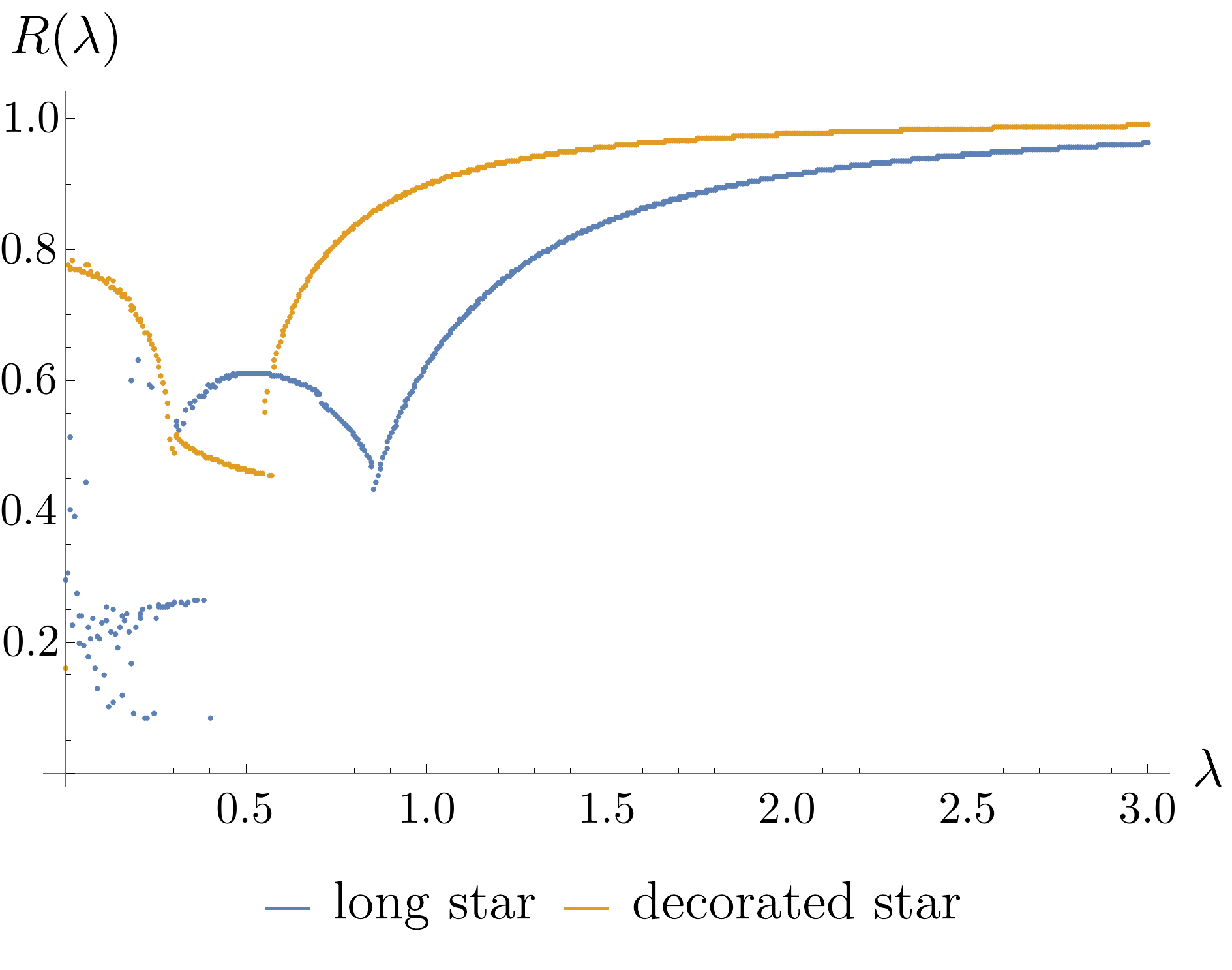} \\ a)
\end{figure}
It was shown that the vertices with high degree significantly affect the synchronization process (see, for instance \cite{radicchi}) hence one can expect that increasing $m$ the dependence $R=R(\lambda)$ for the decorated star graph significantly differs from the order parameter behavior on the long star graph. This observation is consistent with simulations (fig.~\ref{fig:compare_lstar_dstar}b). We conclude that the Kuramoto model on the decorated star graph exhibits behavior similar to the model on long star graph but it is modified by presence of additional bonds $m$ through the ray. For large enough $m$, $m\sim n$, the behavior of order parameter becomes strongly hysteretic.
\begin{figure}
    \centering
    \includegraphics[width=0.8\linewidth]{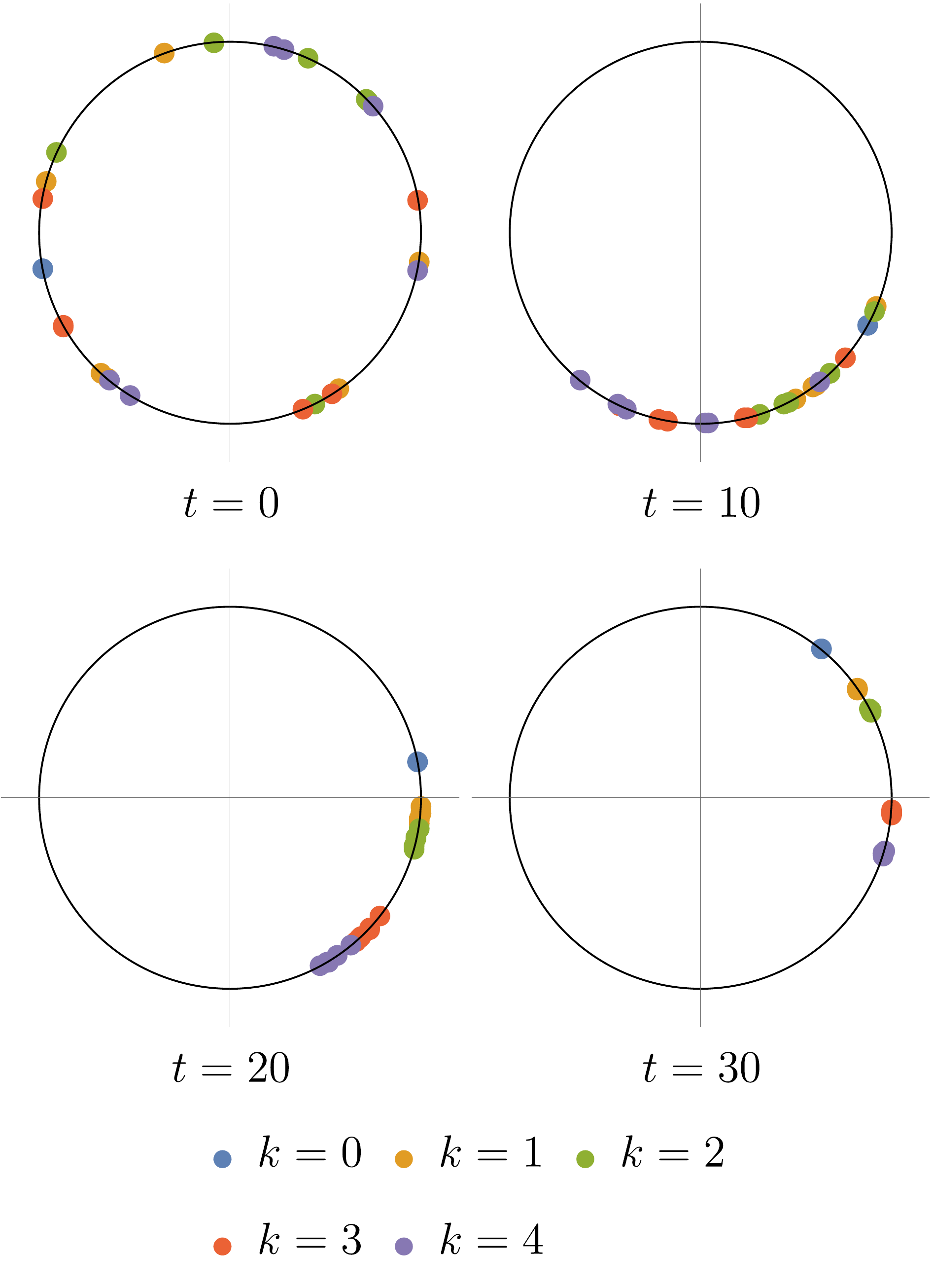}
    \caption{Synchronization dynamics on the decorated long star with $n=6$, $p=4$, $m=3$, the index $k$ denotes the layer, $k=0$ represents the hub}
    \label{fig:dstar_snapshots}
\end{figure}
Finally, the synchronization process on the decorated star graph exhibits step-by-step time behavior as well. It can be captured via the system snapshots at different times, see. fig.~\ref{fig:dstar_snapshots}. As in the case of long star, the synchronization grabs firstly the nearest neighbors of the central node and the leaves become synchronized at the final step. Synchronized state also has a clustered structure due to  phase differences between oscillators on  rays.

\subsection{"Neuron"-like graph}
We also briefly discuss one more modification of star graph, which is the simple model of neuron. The neuron-like graph can be defined by two parameters: the number $n$ of rays of neuron head and the length of neuron tail $p$. 

\begin{figure}
    \centering
    \includegraphics[width=0.8\linewidth]{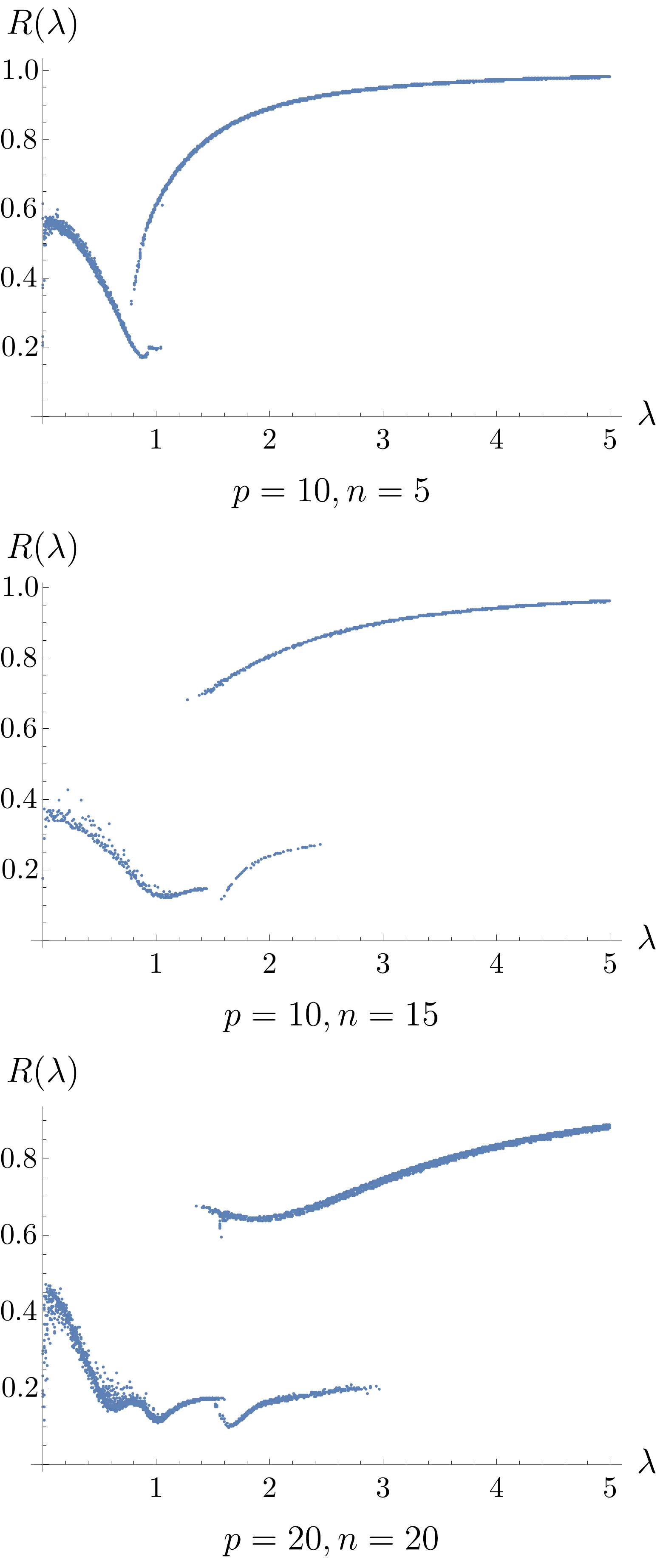}
    \caption{Phase diagram $R=R(\lambda)$ for ``neuron'' graph}
    \label{fig:neuron_order}
\end{figure}

We have performed numerical simulations for different values of $n$ and $p$. The resulting phase diagrams are shown at~\ref{fig:neuron_order}. There are several critical couplings once again with the first order phase transitions.
\begin{figure}
    \centering
    \includegraphics[width=0.8\linewidth]{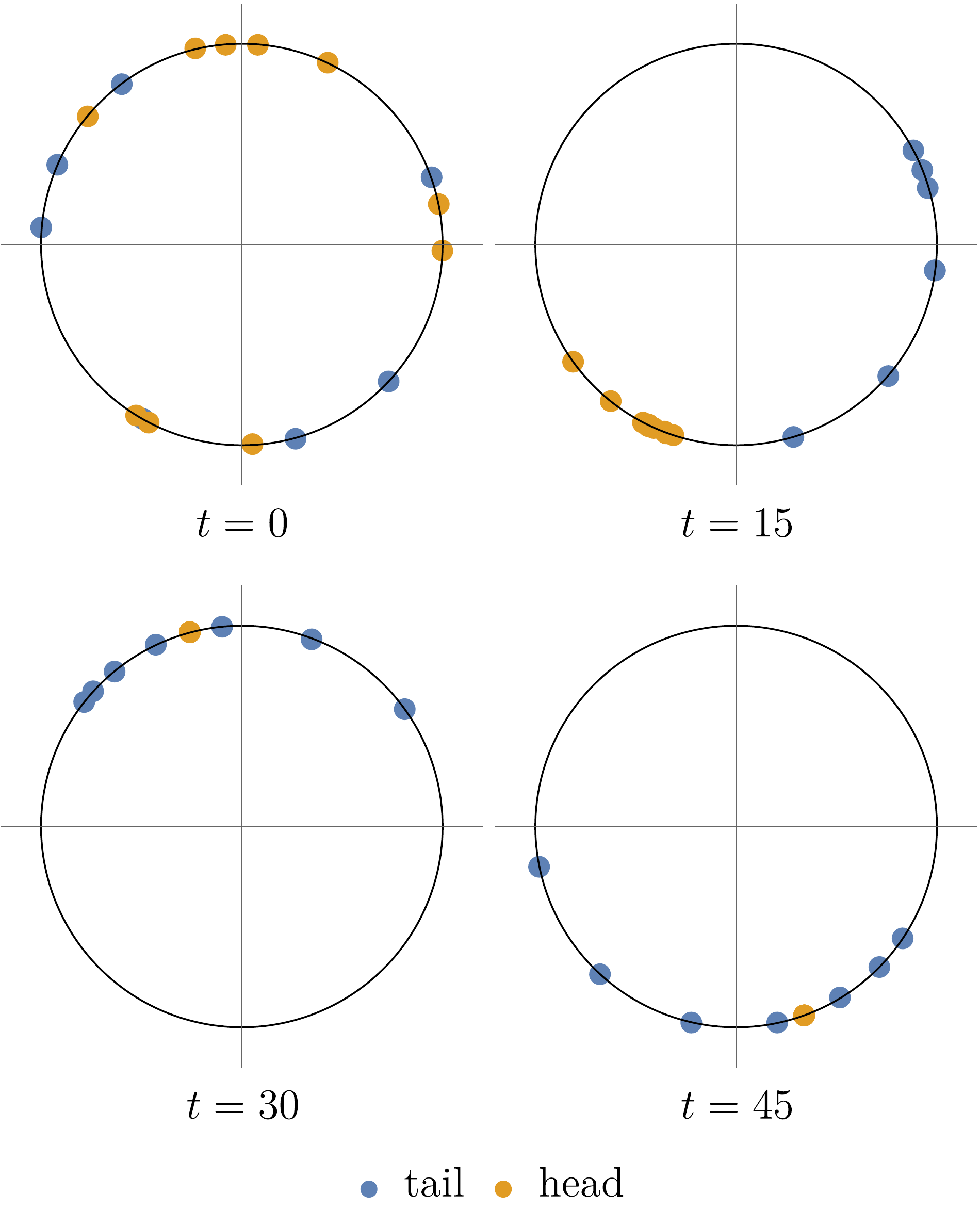}
    \caption{Synchronization dynamics on neuron-like graph with $n=10$, $p=6$.}
    \label{fig:neuron_snapshots}
\end{figure}
The case of ``neuron''-like graph demonstrates step-by-step synchronization in time as well (fig.~\ref{fig:neuron_snapshots}). It starts at the neuron head and spreads through the tail. In the fully synchronized state, there are phase differences in the tail nodes.

\section{Emerging \texorpdfstring{$\mathbb{Z}_p$}{Z(p)} symmetries at strong coupling}
\label{sec:EmergingZSymm}

In this Section we consider the long star graph with large number of rays  $n \gg 1$ \& large length of ray $p \gg 1$ and report a bit counterintuitive phenomenon which  will be shown both analytically and numerically --- there are discrete values of coupling constant when  the Kuramoto order parameter vanishes  in naively completely synchronized state. Somewhat similar splay states were mentioned in \cite{Watanabe1994,Marverl2009,torcini}.

\begin{figure}
    \centering
    \includegraphics[width=0.85\linewidth]{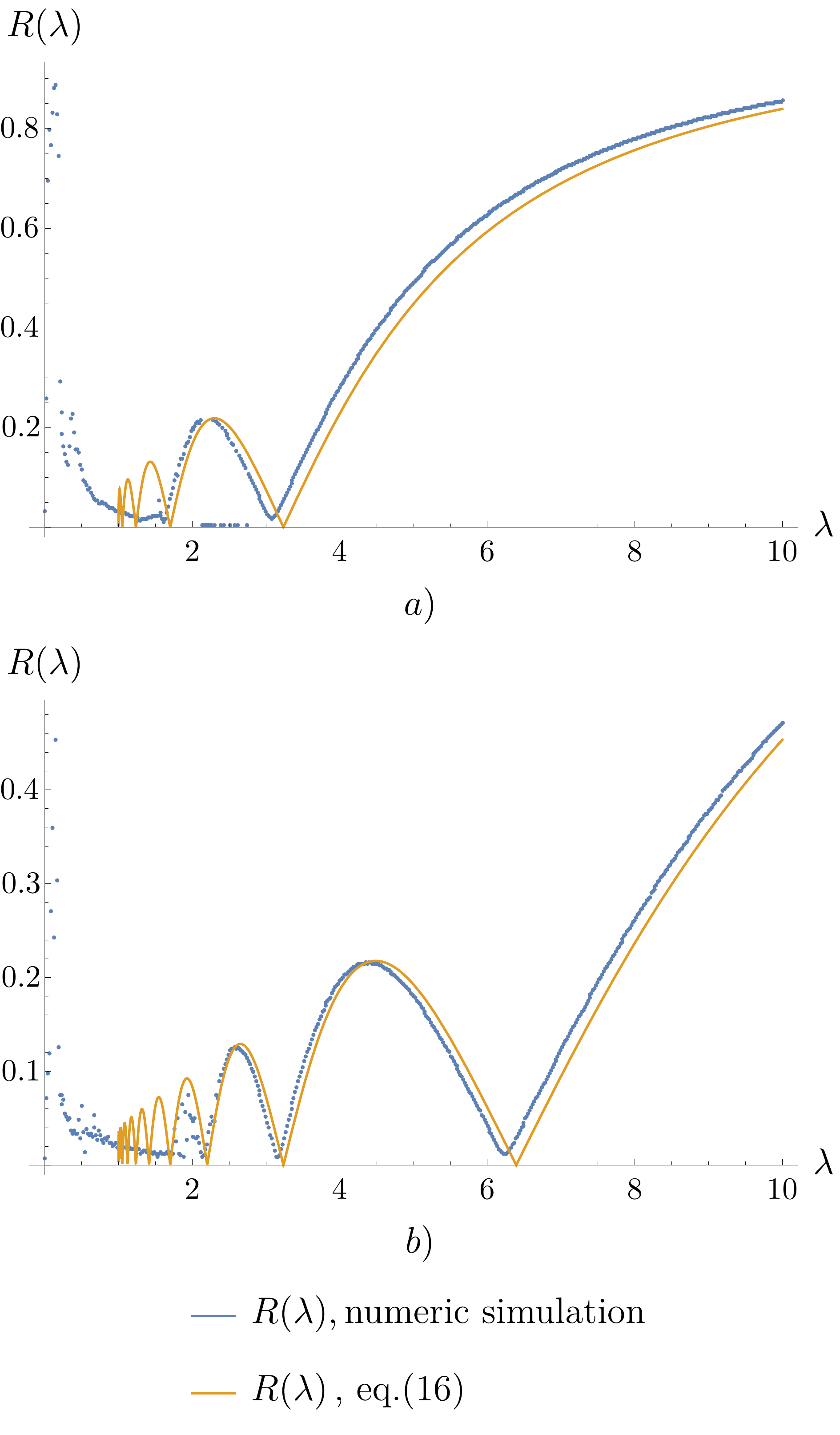}
    \caption{Comparison between order parameter $R(\lambda)$ from numerical simulation for a) $n=p=20$, b) $n=p=40$ long star graphs and analytic result obtained in the limit of large $p$ and large $n$}
    \label{fig:lstar_large_np}
\end{figure}

In synchronized state all rays are equivalent. The solution to dynamical equations is $\theta_k(t)=\Omega t + \varphi_k$, where $\Omega$ is the average frequency and $\varphi_k$ is the constant and $k$ varies from $1$ to $p$. Substituting this expression into  equations , we obtain the system of linear equations for the quantity $\Delta_k=\lambda\sin(\theta_k-\theta_{k-1})$,
\begin{equation}\label{sync_system}
    \begin{gathered}
    \Omega = n + \lambda n\sin(\theta_1 - \theta_0),\\
    \Omega = 2 + \lambda \sin(\theta_k - \theta_{k-1}), k\in [1,p-1], \\
    \Omega =  1 + \lambda \sin(\theta_p - \theta_{p-1}).
    \end{gathered}
\end{equation}
With help of simple algebra, one can found from the system~\eqref{sync_system} the following expression for phase differences on the ray,
\begin{equation}
\label{delta}
    \Delta_k=\frac{2p}{np+1}-1-\frac{2(k-1)}{np+1},
\end{equation}
where $\theta_0$ denotes the phase of the hub and $k$. In the limit of long rays, $p\gg 1$, expanding (\ref{delta}) we obtain
\begin{equation}
    \Delta_k=\left(-1+\frac{2}{n}\right)-\frac{2(kn-n+1)}{n^2p}+\mathcal{O}\left(\frac{1}{p^2}\right).
\end{equation}
From this expansion we also see that for $n \gg 1$ 
\begin{equation}\label{recurr_rel}
   \theta_{k}-\theta_{k-1}=-\arccsc\lambda.
\end{equation}
Setting $k=1$, we obtain $\theta_1=\theta_0 - \arccsc\lambda$ and taking use of  ~\eqref{recurr_rel} one can find all  $\theta_k$. The order parameter can be represented as
\begin{equation}
    r=\frac{e^{i\theta_0}}{np+1}+\frac{n}{np+1}\sum_{k=1}^{p}e^{i\theta_0-ik\arccsc\lambda}.
\end{equation}
In large $p$ and $n$ limit, the contribution from the hub is negligible, so we can safely write
\begin{equation}
    r=\frac{n}{np+1}\sum_{k=1}^{p}e^{i\theta_0(t)-ik\arccsc\lambda}.
\end{equation}
Upon summation of series we get for the order parameter,
\begin{equation}\label{lstar_large_n_order}
|r|^2=\frac{n^2}{(np+1)^2}\frac{\cos(p\arccsc\lambda)-1}{\sqrt{1-(1/\lambda)^2}-1}
\end{equation}
The immediate inspection at $\lambda > 1$ shows that the order parameter has zeros at points $\lambda=\csc(2\pi m/p)$ , $m\in\mathbb{Z}$ and $1\leq m \leq \lfloor p/4\rfloor$. From the eq.~\eqref{lstar_large_n_order} it is clear that the number of zeros is controlled by the value of $p$. Numerical simulation with this limit is in consistent with eq.~\eqref{lstar_large_n_order} which is represented on fig.~\ref{fig:lstar_large_np}. On If one consider limit of large $p$ and finite $n$, the expression for the order parameter can be obtained in the same way but now the hub contribution is not negligible. In such case the expression is more complicated but also demonstrates appearance of $\mathbb{Z}_k$-symmetric structure with some $k$.

\begin{figure}
    \centering
    \includegraphics[width=\linewidth]{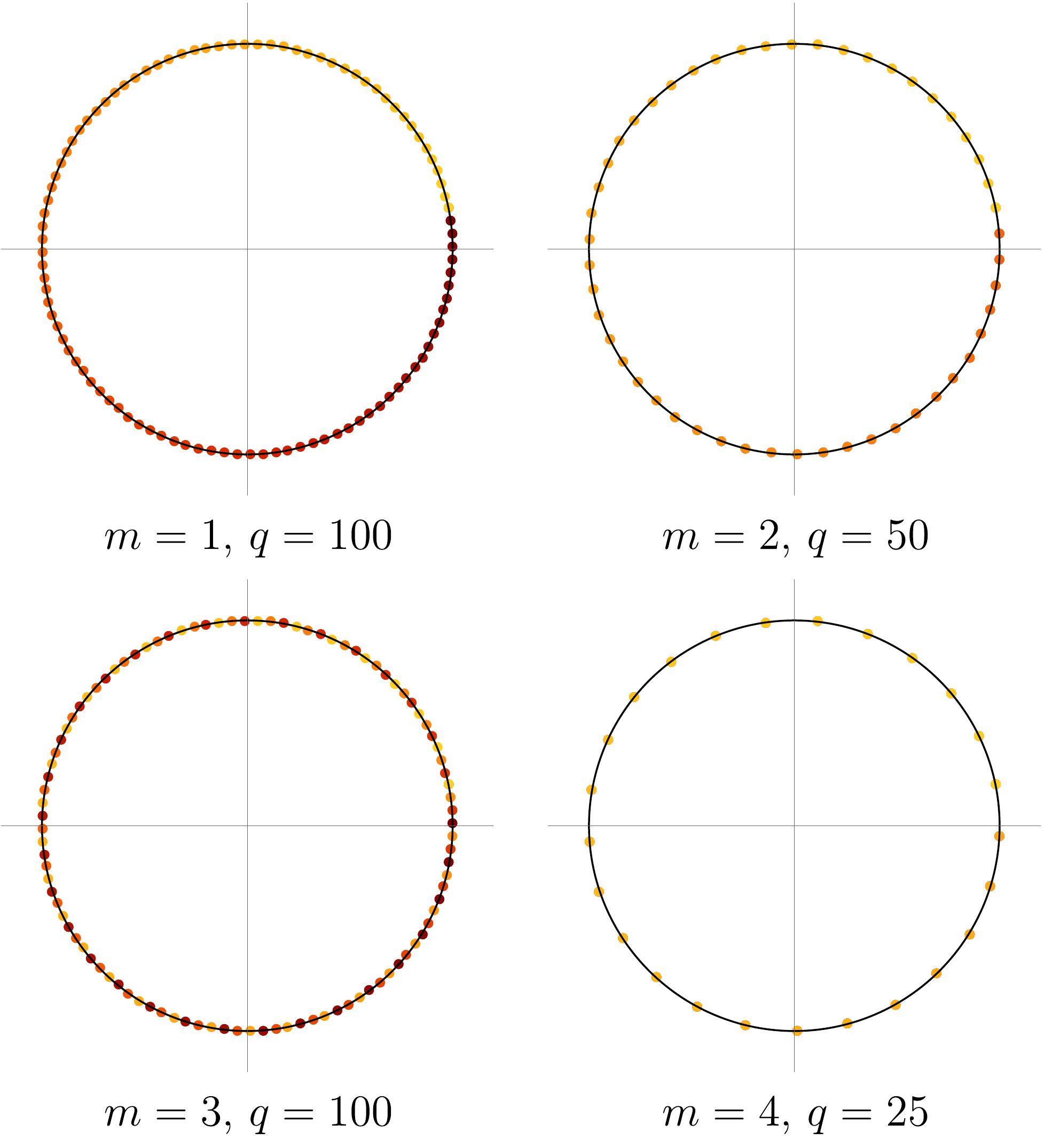}
    \caption{Possible $\mathbb{Z}$-symmetric phases corresponding for different values of $m$ with $p=100$. Brighter colors represent the larger values of $k\in[1,p]$}
    \label{fig:lstar_z_phases}
\end{figure}

\begin{figure}
    \centering
    \includegraphics[width=0.85\linewidth]{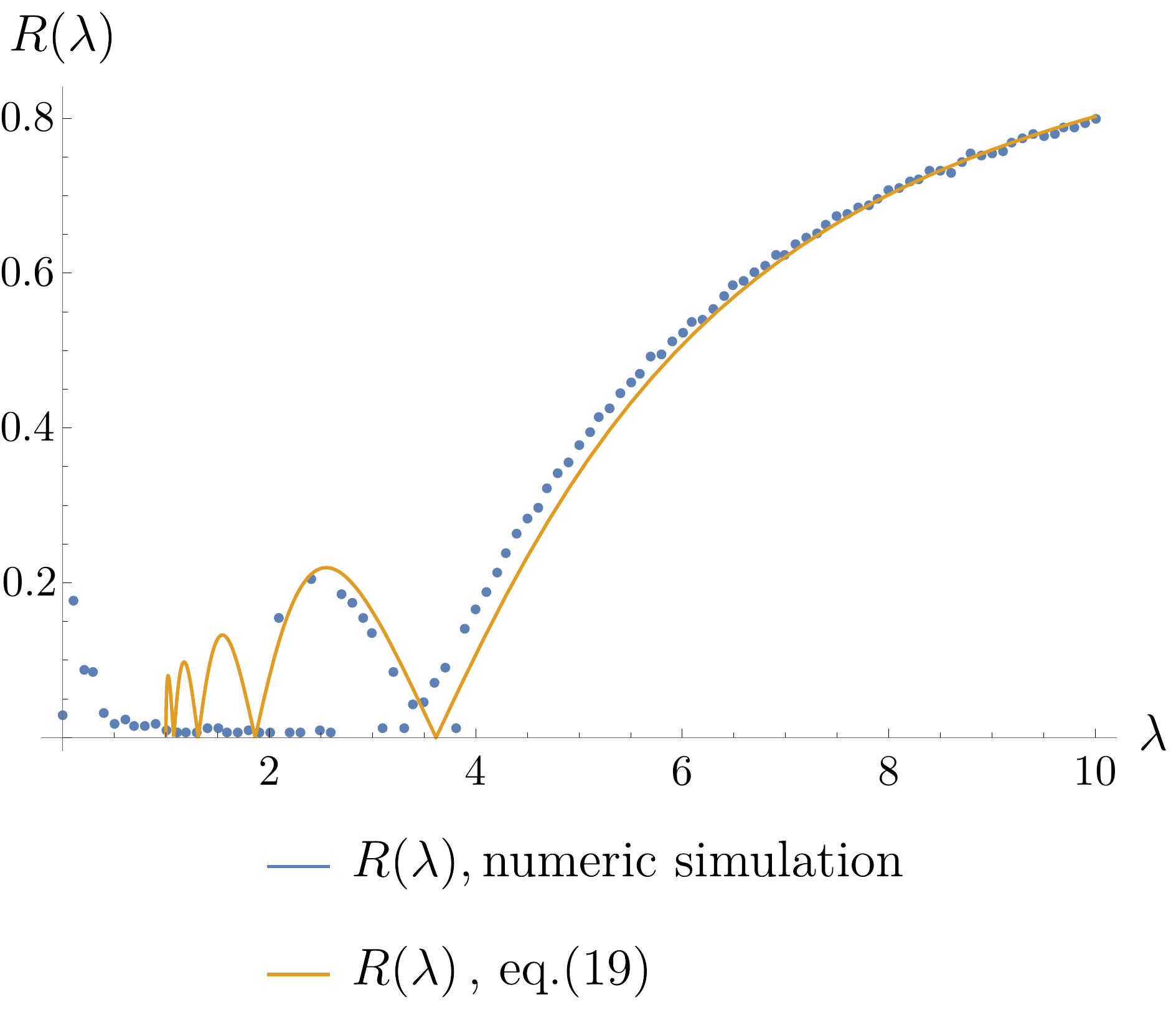}
    \caption{Phase diagram for decorated star with $n=p=30$ and $m=2$}
    \label{fig:dstar_large_n}
\end{figure}

In case of $m=1$ the oscillators form configuration on the unit circle $S^1$ that preserves $\mathbb{Z}_p$-symmetry. For larger values of $m$, this $\mathbb{Z}_p$-symmetric structure reshapes. For $m$ from $1\leq m \leq \lfloor p/4\rfloor$ the appeared structure has $\mathbb{Z}_{q}$ symmetry where $q=p/\text{GCD}(p,m)$. Such configurations have $q$ clusters. Fig.~\ref{fig:lstar_z_phases} represents different possible structures with cyclic group symmetry.
\begin{figure}
    \centering
    \includegraphics[width=0.9\linewidth]{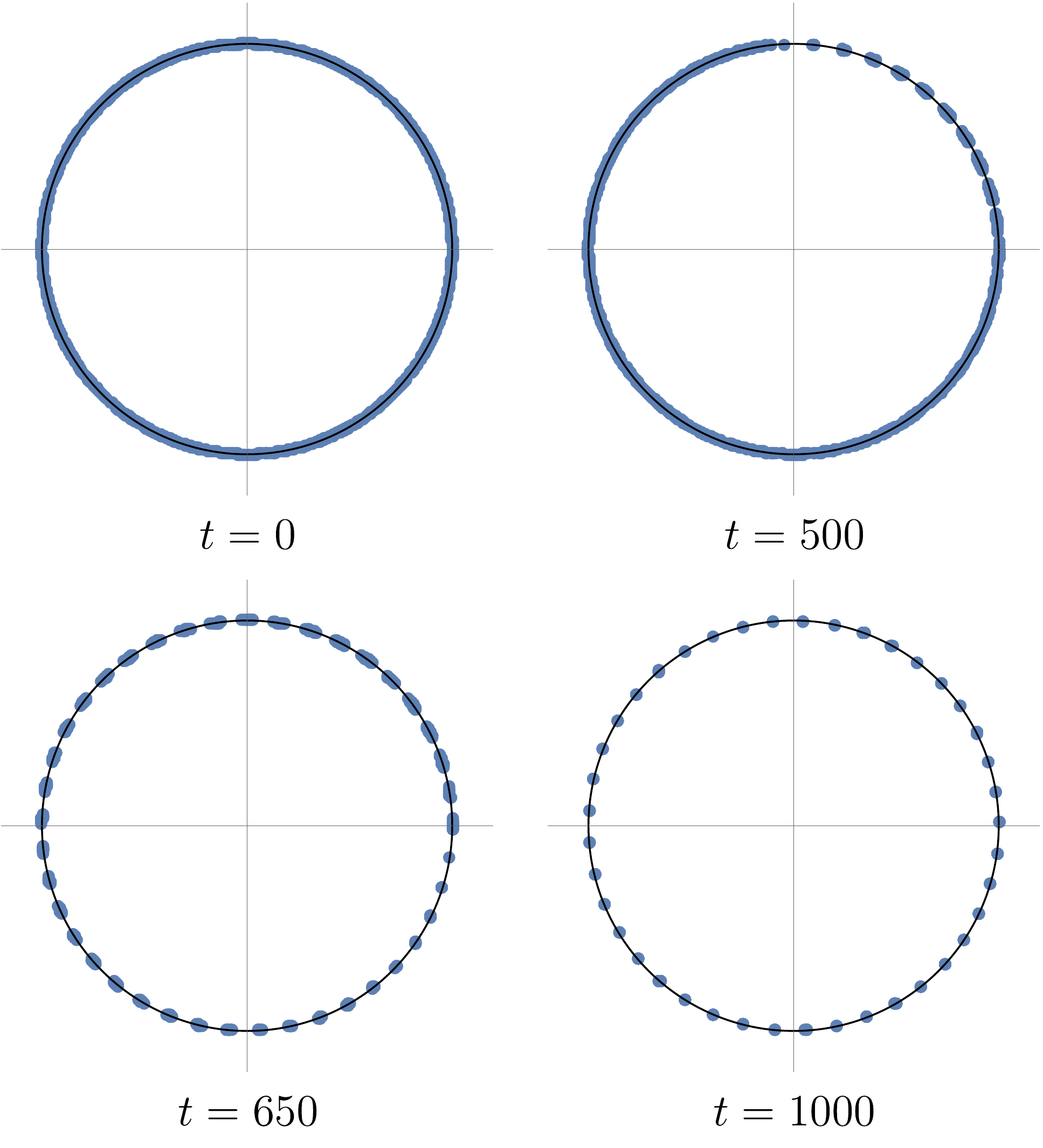}
    \caption{Snapshots of the model on long star graph with $n=p=40$, $R(\lambda=\csc(2\pi/p))\approx 4.4\cdot 10^{-4}$.}
    \label{fig:lstar_large_snapshots}
\end{figure}
What is the origin for the vanishing of the standard Kuramoto order parameter? In the usual Kuramoto model with all-to-all couplings in the synchronized phase the degrees of freedom are grouped together providing $r\neq 0$. For the model on the long star graph the different synchronization patterns occur at discrete points. The particles are organized in several groups in a symmetric manner enjoying the $\mathbb{Z}$-symmetry (it can be captured from the model snapshots with corresponding coupling at large enough times, see fig.~\ref{fig:lstar_large_snapshots}). The system gets synchronized but the Kuramoto order parameter vanishes being the inappropriate order parameter in this phase. 
Similar phenomenon of vanishing of the standard Kuramoto order parameter at discrete values of coupling takes place for the decorated star as well (see fig.~\ref{fig:dstar_large_n}). We get the average degree,
\begin{equation}
    \Omega=\frac{np}{np+1}(m+1).
\end{equation}
which in the limit of large $n$ \& $p$ yields $\Omega\approx m+1$. Then, performing the similar analysis one can find
\begin{equation}
    |r|^2=\frac{4\cos^2(\beta/2)\csc^2[(\alpha+\beta)/2]\sin^2[p(\alpha+\beta)/4]}{p^2}
\end{equation}
where $\alpha = \arccsc\lambda$, $\beta = \arccsc(m\lambda)$. Similarly to the long star case, for decorated long star we get the vanishing order parameter at
\begin{equation}
    \frac{p(\alpha+\beta)}{2}=2\pi k,\quad k\in\mathbb{Z}.
\end{equation}
One can also consider corrections to the mentioned expressions using simple Taylor series expansions.
Note that synchronization of the clusters has been discussed before, see for instance \cite{motter,zaikin}. In this case it was useful to measure the synchronization in terms of order parameter which involves phases of oscillators inside each cluster.

\section{Discussion}
\label{sec:Discussion}

In this short note we have discussed synchronization on  star-like graphs involving long rays: long stars, decorated long stars and "neuron"-like graph. It was shown numerically that in all three cases different ingredients get synchronized at the different critical couplings. The chain of first order phase transitions with clear-cut hysteresis has been observed and all  critical couplings have been found for the long star combining numerical and analytic tools. The interesting new phenomenon concerns the emerging  cyclic symmetry  at discrete values of the coupling constant. These 
fully synchronized states enjoy the   $\mathbb{Z}_{k}$ symmetry with some k  and the conventional order parameter vanishes. The phenomenon takes place for long star and decorated long star and there is qualitative agreement between the analytic results at large $N$ and the numerical simulations. Since the naive order parameter is not suitable in this case it is useful to apply the order parameters invented for the clusterized states in Kuramoto model \cite{motter,zaikin}. 

The Kuramoto model provides some insights for the behavior of the BES condensate of charged degrees of freedom on the arrays of Josephson junctions and the synchronized phase corresponds to the formation of the condensate
on the particular graph. The multistep synchronization implies that in some interval of the couplings the BEC is inhomogeneous in the radial direction from the hub and we have some number of synchronized clusters and 
desynchronized degrees simultaneously. In the context of synchronization it is some version of the chimera state while in the context of the BCS superconductivity it can be considered as the version of pseudogap phase when only part of the Cooper pairs gets synchronized. 

The phase with $\mathbb{Z}_{p}$ symmetry is especially interesting from the Josephson arrays viewpoint. In this case at strong coupling  we see that there are synchronized clusters at $\theta_k=2\pi k/p$ in rotating frame. Each cluster corresponds to the union of oscillators at distance $k$ from the hub therefore at each ray we have the phase wave as a ground state at these discrete couplings. BEC at each ray is non-vanishing and modulated.

There are several interesting questions for the future study. First, we did not use the power  of the  M\"{o}bius group naturally acting at the set of points on $S^1$ however it could be expected that for the symmetrically enough graphs the synchronization can be related  to the particular flows on the  M\"{o}bius group generalizing the corresponding analysis for the full graph and star graph. Another question concerns the effects of disorder. Usually disorder is added via the distribution of the internal frequencies or just by adding a random term into the Kuramoto equation. However it is quite natural to consider the Kuramoto model on  exponential random graph which is the discrete analogue of the matrix model for a orthogonal ensemble. In this case  we would deal with the Kuramoto model on the fluctuating geometry which could have  the phase transition itself. The investigation of the phase field on the random graph is the discrete version of the matter interacting with 2D gravity and the effects of the random geometry on the condensate formation certainly are of  interest. It would be also interesting to investigate higher-order Kuramoto model discussed in \cite{bianconi} on fluctuating simplicial complexes.

The somewhat related question concerns a appearance of the version  of the Kuramoto model on the full graph in the system of complex SYK model flavored with the Hubbard interaction \cite{Gorsky2020} which induces the superconductivity. It was argued  that at large enough Hubbard coupling the quantum Kuramoto model on the full graph has been generated and both pseudogap and BCS phases have been observed. It would be interesting to perform the similar analysis for the complex SYK+Hubbard model on the star-like graph. It would be also interesting to relate our findings with the desynchronization in the BCS condensate formation which takes place in some small interval of couplings when the coupling constant get increased \cite{Barankov2006}.

The work was supported by  Russian Science Foundation grant 21-11-00215.
		
\bibliography{references}

\end{document}